\documentclass[10pt,a4paper]{spie}
\usepackage[utf8]{inputenc}
\usepackage[dvipdf]{graphicx}
\usepackage{makeidx}
\usepackage{color}
\usepackage{multicol}

\title{
    Automatic unit for measuring refractive index of air based on Ciddor equation and its verification using direct interferometric measurement method
    }

\author{
	V. Hucl,
    M. Cizek,
	J. Hrabina,
	B. Mikel,
	S. Rerucha,
	Z. buchta,
	P. Jedlicka,
	A. Lesundak,
	J. Oulehla,
	L. Mrna,
	M. Sarbort,
	R. Smid,
	J. Lazar,
	O. Cip
\skiplinehalf
Institute of Scientific Instruments of the Czech Academy of Sciences (ISI),\\
 Kr\'{a}lovopolsk\'{a} 147, 612 64 Brno, Czech Republic
\skiplinehalf
}

\authorinfo{Further author information: please send correspondence to e-mail address treak@isibrno.cz (V. Hucl). arXiv rendition and further development by S. Rerucha res@isibrno.cz}

\begin{document}
\maketitle


\begin{abstract}

In scanning probe microscopy laser interferometers are usually used for measuring the position of the probe tip with a metrological traceability. As the most of the AFM setups are designed to work under standard atmospheric conditions the changes of the refractive index of air have an influence to measured values of the length with $1.0 \times 10^{-4}$ relatively. In order to achieve better accuracies the refractive index of air has to be monitored continuously and its instantaneous value has to be used for compensating the lengths measured by all of the interferometric axes. In the presented work we developed a new concept of an electronic unit which is able to monitor the refractive index of air on basis of measurement of ambient atmospheric conditions: temperature, humidity, pressure of the air and the CO2 concentration. The data processing is based on Ciddor equation for calculating the refractive index of air. The important advantage of the unit is a very low power consumption of the electronics so the unit causes only negligible temperature effects to the measured environment. The accuracy of the indirect measuring method employed by the unit was verified. We tested the accuracy in comparison with a direct method of measuring refractive index of air based on an evacuatable cell placed at the measuring arm of a laser interferometer. An experimental setup used for verification is presented together with a set of measurements describing the performance. The resulting accuracy of the electronic unit falls to the $4.1 \times 10^{-7}$ relatively.

\end{abstract}

\\
\\
{\bf DOI:} 10.1117/12.2020756 \\

Note: This is a preprint rendition of a conference paper V. Hucl et al,  Automatic unit for measuring refractive index of air based on Ciddor equation and its verification using direct interferometric measurement method, Proc. SPIE, vol. 8788, pp. 878837, 2013.

%
%
%
%


\section{Introduction}

The precise measurement of lengths in the industrial or laboratory conditions by laser interferometers relies very much on the knowledge of a value of the refractive index of air. The particles and the air itself make the measuring optical path longer than the geometrical is. Therefore the determination of the index value is necessary to measure in real-time when the measurement by lasers is carried-out. For laboratory measurement is very useful a direct measurement of the index by a laser refractometer. This measuring setup works as an interferometer where fixed known geometrical length is detected by laser interferometer. The fluctuation of the index is then visible as a change of the optical path length at the output of the interferometer [1]. On the other hand the direct method needs very complicated optical setup. In this case there is also problem with placing of the refractometer setup close to the measuring interferometer used for the desired distance detection.


\section{Methodology}
\label{met}

For simple and relatively precise measurement of the refractive index of air the non-direct measurement is very useful. This principle is based on measurement of set of quantities of the ambient air which cover the measuring area: temperature, relative humidity and air pressure. From known values of these quantities we determine this index on basis of one of these empiric formulas: Edlen [2], Bonsch and Potulski [3], or Fira [4]. In general all of above mentioned formulas came from fundamental equation founded by Edlen [2]. The empirical measurement and comparison of calculated values with direct measurement of the index by interferometric way, i.e. laser refractometers [5,6] led to additional improvement. This modified Edlen formula can be expressed by the following procedure:

\begin{eqnarray*}
A &=& 8342.54, B = 2406147, C = 15998, D = 96095.43, E = 0.601, F = 0.00972, G = 0.003661 \\
S &=& \frac{1}{\lambda^2}\\
n_s &=& 1 + 10^{-8}\left[A + \frac{B}{130-s} + \frac{C}{38.9-S}\right]\\
X &=& \frac{1 +  10^{-8} (E - Ft)p}{1+ Gt} \\
n_{tp} &=& 1 + \frac{p(n_s-1)X}{D}\\
n_s &=& n_{tp} - 10^{-10}\frac{292.75}{t+273.15}(3.7345-0.0401S)p_v
\end{eqnarray*}

where t is the temperature of air [$\deg$C],  $p$ is the pressure [Pa], $p_v$ is the water vapor partial pressure, $\lambda$ is the wavelength [$\mu$m]. Water vapor partial pressure we calculate using $p_v = \frac{RH}{100}p_{sv}(t)$ expression, where $RH$ is the relative humidity [\%] and $p_{sv}$ is the saturation vapor pressure [Pa].

On basis of the research of the relation between the refractive index and a carbon dioxide concentration the Ciddor formula is nowadays considered as a very reliable equation [7]. When using the Ciddor formula it is necessary to express humidity as a mole fraction:
\begin{eqnarray*}
	\alpha &=& 1.00062, \beta = 3.14e-8, \gamma =5.60e-7\\
	f(p,t) &=& \alpha + \beta p + \gamma t^2\\
	x_v &=& \frac{RH}{100} f(p,t) \frac{p_{sv}}{p}
\end{eqnarray*}

 For calculation the refractive index of air  we define first following constants:
 \vskip .3cm
 
\begin{tabular}{l l l l}
$w_0=295.235$ &  $w_1=2.6422$ & $w_2=-0.03238$ & $w_3=0.004028$ \\
$k_0=238.0185$ & $k_1=5792105$ & $k_2=57.362$ & $k_3=167917$ \\
$a_0=1.58123e-6$ & $a_1=-2.9331e-8$ & $a_2=1.1043e-10$ & \\
$b_0=5.707e-6$ & $b_1=-2.051e-8$ & & \\
$c_0=1.9898e-4$ & $c_1=-2.376e-6$ & $d=1.83e-11$ & $e=-0.765e-8$ \\
$p_{R1}=101325$ & $T_{R1}=288.15$ & $Z_a=0.9995922115$ &  \\
$\rho_{vs} = 0.00985938$ & $R=8.314472$ & $M_v=0.018015$ & \\
\end{tabular}

Then we insert measured atmospheric parameters to equations:

\begin{eqnarray*}
	r_{as} &=& 1e-8\left(\frac{k_1}{k_0-S}+\frac{k_3}{k_2-S}\right)\\
	r_{vs} &=& 1.022-8 (w_0+w_1S+w_2S^2 +w_3S^3)\\
	M_a &=& 0.0289635+1.2011e-8(x_{CO2}-400)M\\
	r_{axs} &=& r_{as}(1+5.34e-7(x_{CO2}-450))\\
	Z_m &=& 1 - \frac{p}{t + 273.15} (a_0 + a_1t + a_2t^2 + (b_0+b_1t)x_v + (c_0 + c_1t)x^2_v) + \frac{p}{t + 273.15} (d + e*x^2_v)\\
	\rho_{axs} &=& \frac{p_{R1}M_a}{Z_a R T_{R}}\\
	\rho_{v} &=& \frac{x_v p M_v}{Z_m R (t + 273.15)}\\
	\rho_{a} &=& \frac{(1-x_v) p M_a}{Z_m R (t + 273.15)}\\
	n &=& 1 + \frac{\rho_a}{\rho_axs}r_{axs} + \frac{\rho_v}{\rho_vs}r_{vs}
\end{eqnarray*}

\noindent
where $t$ is the temperature of air [$\deg$C],  $p$ is the pressure [Pa], $\lambda$ is the wavelength [$\mu$m], RH is the relative humidity [\%], $x_{CO2}$ is the CO2 concentration [ppm] and $p_{sv}$ is the saturation vapor pressure [Pa].
The Ciddor formula is more complicated then the modified Edlen formula but it produces better precision of measurement of the refractive index.


\section{Experimental setup}

\begin{figure}[htbp]
	\centering
	\includegraphics[width=.8\textwidth]{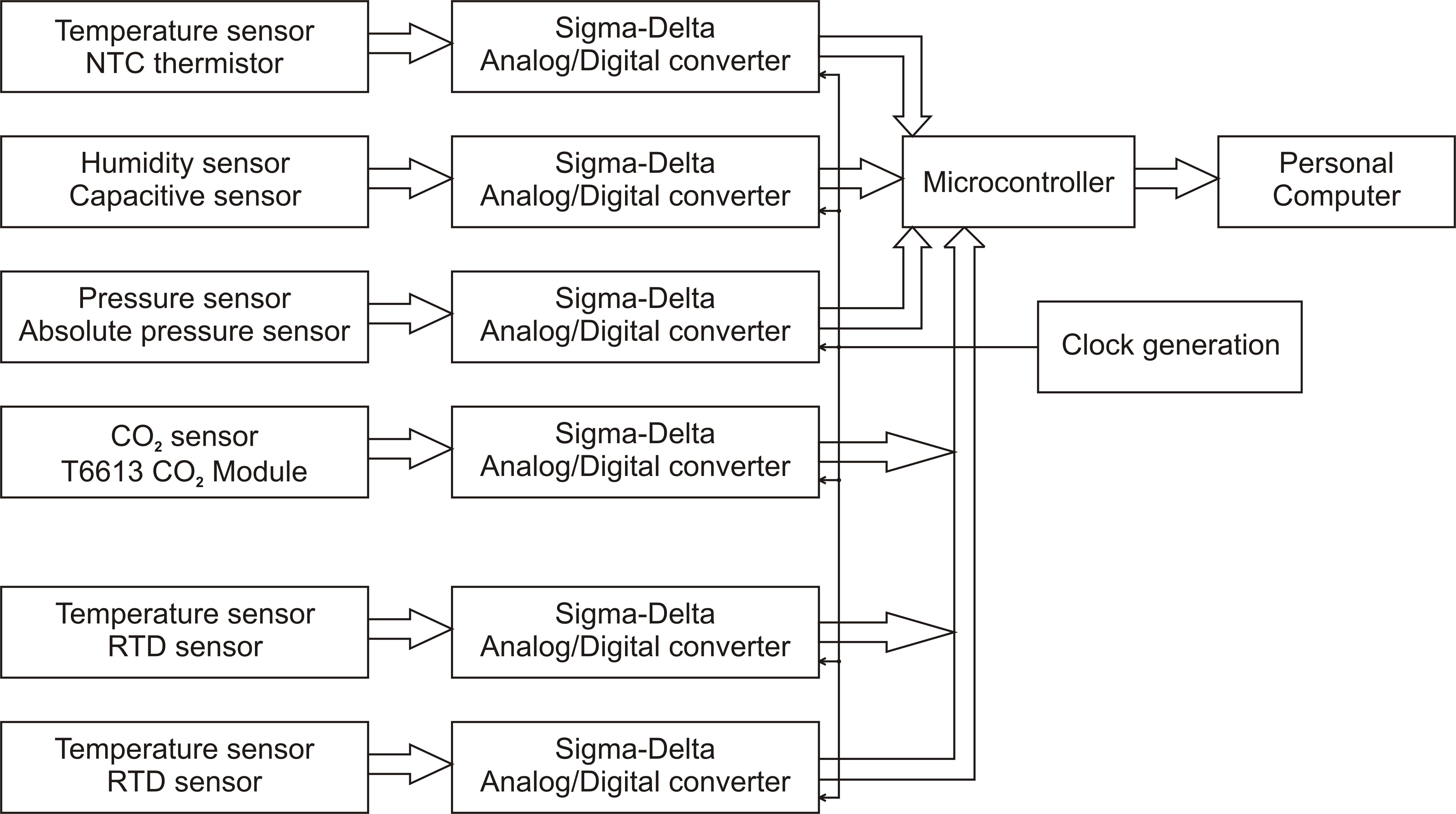}
	\caption{Schematic diagram of the electronic unit for precise measurement of the refractive index of air. The unit includes separated 6 AD converters which sample the voltage signals from different sensor for monitoring of atmosphere conditions. Connected microcontroller is the heart of the unit which processes all of sampled signal (calibration, refractive index formulas calculation and transfer to the personal computer). The sampling rate is controlled by a separated oscillator (clock generation unit).}
	\label{f1}
\end{figure}

On the basis of the modified Edlen or Ciddor formula is possible to put together the measuring unit which has implemented a set of precise sensors, signal processing and final calculation of the refractive index of air in the real-time regime. The main goal of the presented work was the design of the electronic unit which is able to measure the refractive index of air with high accuracy and in real-time. The necessary task of the unit is also in the fast transfer of the measured values into the central electronics where data from the distance measuring interferometer system is processed. The schematic diagram of the unit for non-direct precise measurement of the refractive index of air is viewed in Fig. \ref{f1}.

The main core of the unit uses the single-chip microcontroller Freescale 68HC08 [8], which is responsible to collecting data from sensors, calculation of the refractive index of air and transmitting measured and calculated data to PC. All of sensors are connected to the unit by a set of separated 16-bit Sigma/Delta Analog/Digital converters Analog Devices AD7715 [9]. This set of converters is chosen due to the requirement of very high resolution of each quantity and to prevent a possible noise coming from sensors and random electromagnetic spikes. The single-chip microcontroller does also a communication with PC through a serial communication bus. In that case we use Controller Area Network bus (CAN) [10] with CANopen communication protocol [11].

The view of the completed electronics is in Fig. 2. The sensor Humirel HTM2500 [12] is used to temperature and humidity measuring, which have integrated both sensors in one package. The temperature sensor is NTC thermistor type and measured temperature is evaluated from resistivity change on basis of calibration curve. The humidity sensor is capacitive one with special solid polymer structure, which has a high protection to chemicals and has the fast response time. The humidity value corresponds to the value of voltage at the output of the sensor. The pressure is measured by Motorola MPXH6101A [13] piezoresistive transducer, which is based on silicon pressure probe. The air pressure corresponds with the value of voltage at the output as well as at the humidity sensor.

\begin{figure}[htbp]
	\centering
	\includegraphics[width=.8\textwidth]{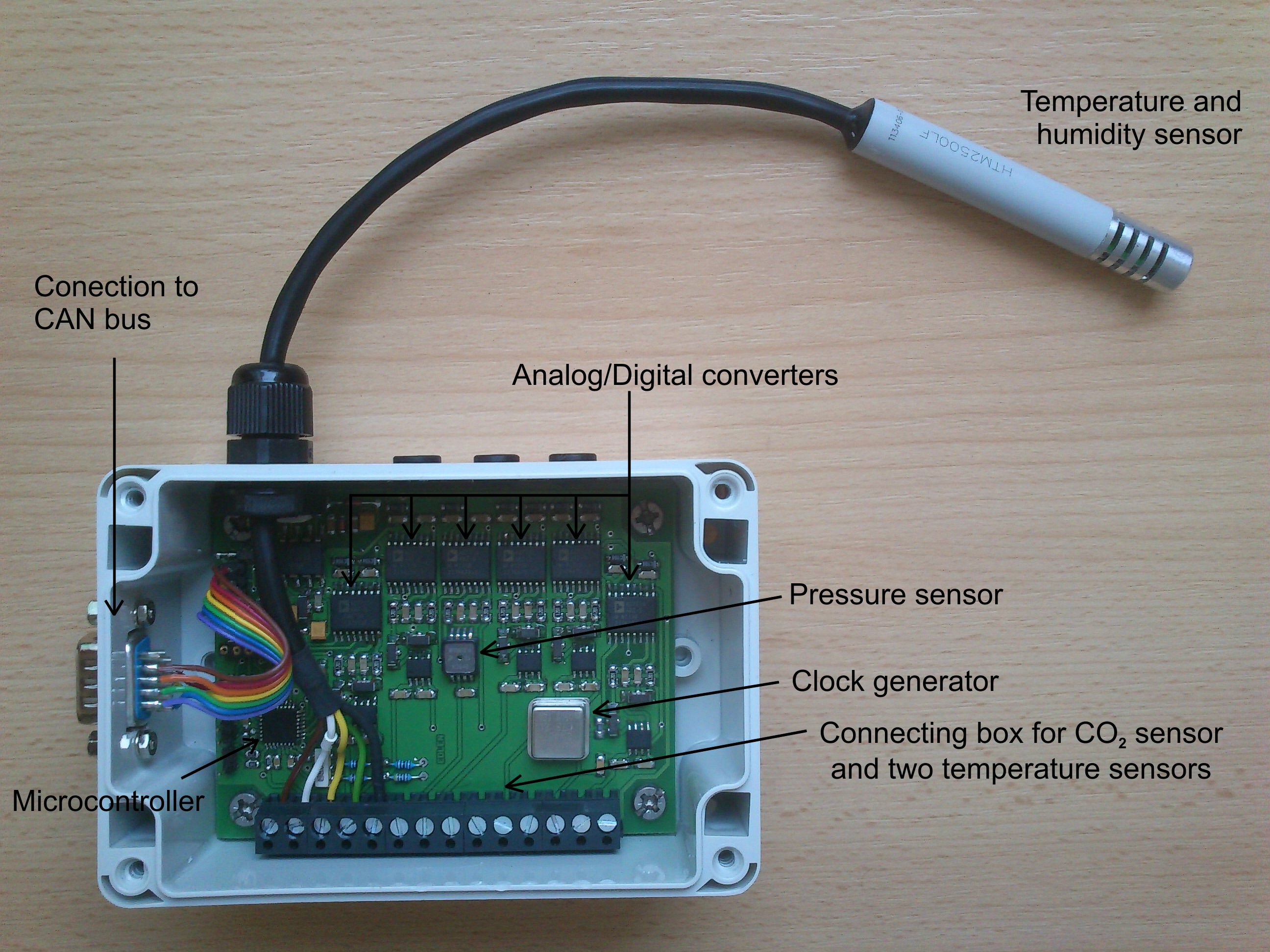}
	
	\caption{The view of the final sample of the electronic unit for precise measurement of the refractive index of air. }
	\label{f2}
\end{figure}

We designed the electronic unit for connecting of two additional temperature sensors which can be used to monitoring of the temperature in crucial part of the experimental setup with the distance measuring interferometer. They are situated for temperature measurement of the frame and some necessary parts of the measured element. We adapted it for connecting of Analog Devices AD22100 sensors [14], those are ratiometric temperature integrated circuits whose output voltages are proportional to the temperature in linear dependence. We also equipped the electronics for linking of the external CO2 sensor Telaire 6613 CO2 Module, those measure method is based on the non dispersive infrared (NDIR) with gold plated optics and on a diffusion sampling. So the output voltage of this sensor is sampled by the appropriate AD converter AD7715 as was mentioned before.

The primary sampling rate of the set of AD converters is 50 samples/second. This higher rate is used for improvement of the signal to noise ratio by the microcontroller where digital low-pass filtering with 0.5 Hz frequency is implemented for each measured channel. Then the down-conversion for each channel with decimation is applied so the output sampling rate runs about 5 samples/second.

The microcontroller of the electronic unit is responsible also for polynomial correction of all of measured signals.  The calibration constants are stored in the flash memory of the controller. When the input samples are processed by filters then the correction by these constants is provided. The final part of the digital processing covers the calculation of the modified Edlen and Ciddor formula. The corrected samples from sensors and calculated values of the refractive index are then transferred to the main computer. There is implemented a control and visualization software where all of values are presented as instantaneous numbers and as the graphical recordings in the time.

\begin{figure}[htbp]
	\centering
	\includegraphics[width=.8\textwidth]{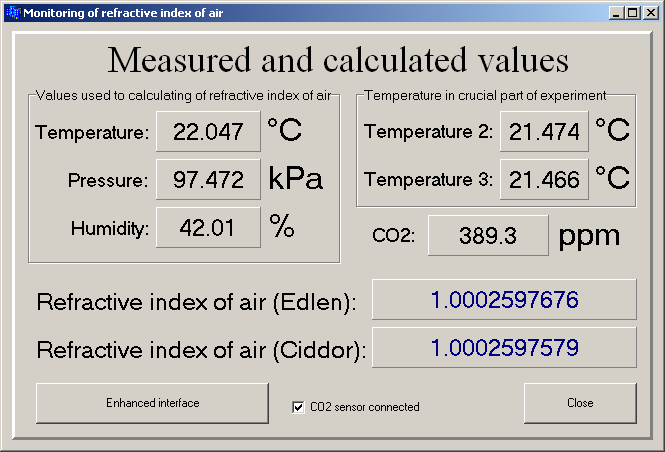}
	
	\caption{User interface to display of measured and calculated values by the electronic unit}
	\label{f3}
\end{figure}

The unit is designed with the stress to minimizing the power losses in the measuring area. All of circuits are chosen as a low-power. There are no voltage stabilization circuits or other heat producing devices. All of these parts are separated to the supplying card inserted into the main electronic rack where detection of interferometric values is also provided. The print-screen of the graphic user interface is in Fig. 3, where instantaneous value of the refractive index is visible. We implemented calculation of the refractive index by modified Edlen and Ciddor formula at the same time.


\section{Experimental results}

Before installation of the unit we provided calibration of all measuring sensors. We used etalon sensors from Vaisala company [15] as the reference. Then we put together experimental testing frame where this Vaisala measuring unit for humidity, pressure and ambient temperature is used. For calibration of the CO2 sensor unit we used precise carbon dioxide gauge probe FYA 600 CO2 and detection instrument ALMEMO 1030-2 from company Ahlborn, Germany [16].
After the calibration process of all of sensors we continued with verification of the refractive index measuring capability with our laboratory laser refractometer. We installed all of sensors of the unit to the refractometer measuring area. Then we provided long-time measurement of the refractive index of air by both methods: refractometer (direct) and electronic unit (non-direct). During this measurement the unit generated the calculated values based on Edlen and on Ciddor formula at the same time. We collected all of measured data with PC computer. The resulted values of all atmospheric quantities are shown in Fig. 4.

\begin{figure}[htbp]
	\centering
	\includegraphics[width=.8\textwidth]{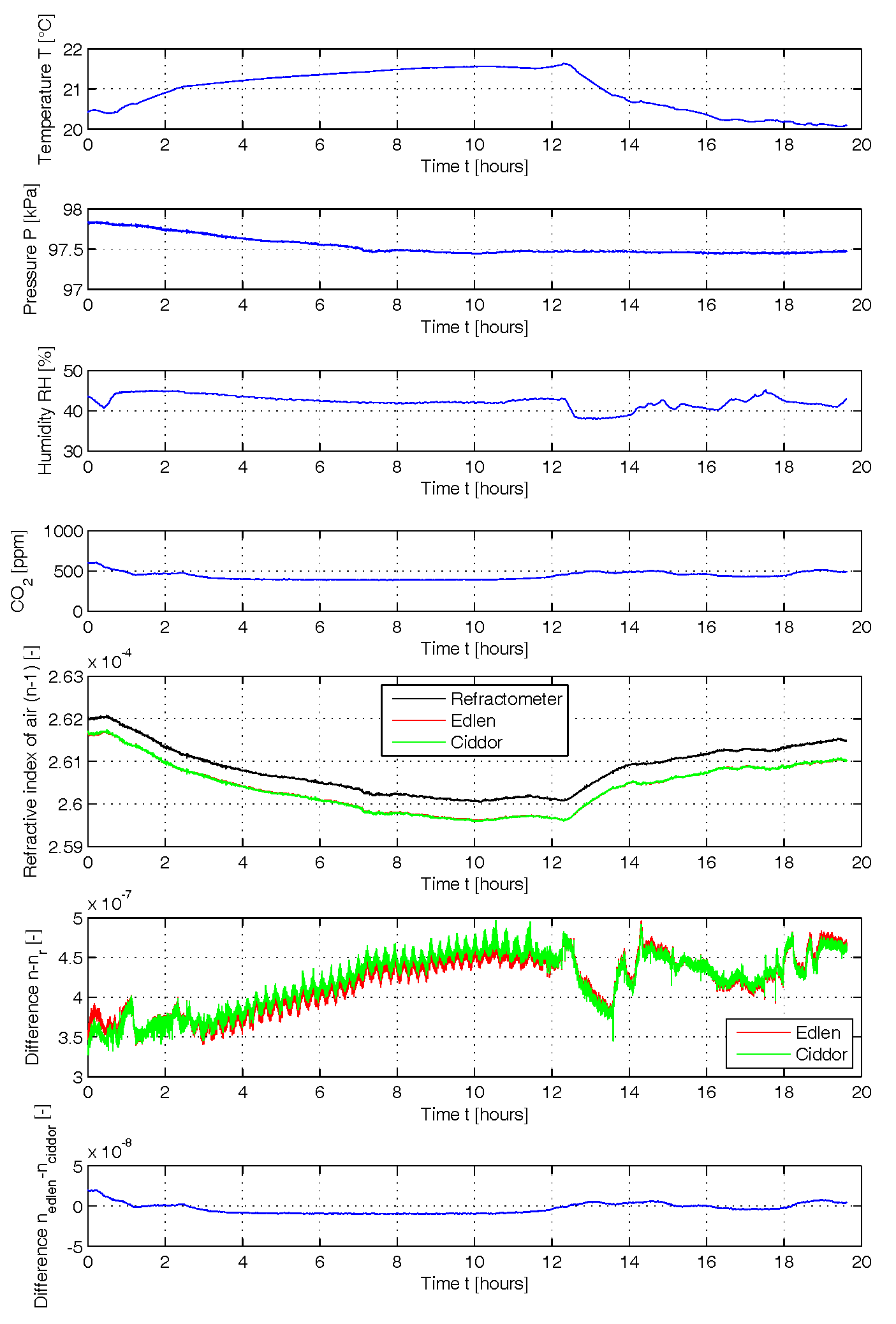}
	\caption{Values of all of atmospheric quantities measured by the electronic unit. These records show forced changes of some values by changing of parameters of the air condition unit installed in the laboratory}
	\label{f4}
\end{figure}

During these 20 hours recordings we controlled some parameters of the air condition unit. We changed temperature and the relative humidity in the laboratory. We also asked 3 people to stay in the laboratory for few hours close to the end of the measuring period. As is visible in the record of carbon dioxide concentration CO2 sensor detected very reliable the increasing of this quantity. The stay of these persons also made influence to relative humidity thanks to their breathing. The humidity sensor also detected fluctuation of this quantity reliable.

The comparison of calculated values of modified Edlen and Ciddor formulas are also presented in Fig. 4. There is viewed also the record of the refractive index measured by the laser refractometer. The small deviation between non-direct and direct measurement is at $4.1 \times 10^{-7}$ level as an average for this 20 hours record. This possible deviation can cause 0.4 nm error for 1 mm measured distance or i.e. 400 nm for 1 m distance. The interesting record is also placed at the bottom of Fig. 4. There is deviation between modified Edlen and Ciddor formulas. With the respect of the carbon dioxide concentration there is strictly visible the influence of this quantity to deviation between these two formulas. But the standard deviation of the refractive index difference is only in the $1.2 \times 10^{-8}$ level what causes errors about 12 nm in the distance of 1 m. But for nanometrology application as the AFM microscopy is Ciddor calculation very necessary. The fluctuation of the index is in this situation comparable with a noise floor of used lasers which drives the measuring interferometer [17].


\section{Conclusion}

The presented work showed the principle of measurement of the refractive index of air by the non-direct method. We described block scheme of the electronic unit which is very useful for measurement of the index in scanning probe microscopes or for measurement in the industrial areas. We implemented modified Edlen formula and Ciddor formula calculation of the refractive index. After the calibration process we made the comparison measurement between the electronic unit and the laser refractometer with evacuatable cell. We observed deviation of the refractive index measurement in $4.1 \times 10^{-7}$ level. We made the analysis of the influence of the carbon dioxide concentration to the both formulas calculation. The unit is eligible to use with precise scanning probe microscopes equipped with multiaxes interferometric measurement system or as the correction unit for industrial interferometers.


\section*{REFERENCES}
\begin{itemize}
\item[[ 1]]		Číp O, Petrů F, Matoušek V and Lazar J: Direct measurement of index of refraction of air by means of high-resolution laser interferometry Physica Scripta T118 48-50, 2005.
\item[[ 2]]		Edlén, B.: The refractive index of air. Metrologia, 2, 71, 1966.
\item[[ 3]]		Bönsch, G., Potulski, E. Measurement of the refractive index of air and comparison with modified Edlén’s formulae. Meterologia, 35, 133-139, 1998.
\item[[ 4]]		Fíra, R.: The refractive index of air. Fine mechanics and optics, 41, 218-221, 1996.
\item[[ 5]]		Yacoot A, Koenders L: Aspects of scanning force microscope probes and their effects on dimensional measurement, 41, Article Number 103001, 2008.
\item[[ 6]]		Číp O, Petrů F, Matoušek V and Lazar J: Direct measurement of index of refraction of air by means of high-resolution laser interferometry Physica Scripta T118 48-50, 2005.
\item[[ 7]]		Ciddor PE, Hill RJ: Refractive Index of Air, Applied Optics, 38, 1663-1667,1999.
\item[[ 8]]		Freescale Inc.: Family manual MC68HC08 microcontroller, Data Sheet, 2006.
\item[[ 9]]		Analog Devices, Inc.: AD7715 – 16-bit, Sigma-Delta ADC, Data Sheet, 1998.
\item[[ 10]]	Standard ISO 11898-2: Controller Area Network.
\item[[ 11]]	Standard EN 50325-4: CAN-based higher-layer protocol for embedded control system.
\item[[ 12]]	Humirel, Inc.: HTM2500 Relative humidity/temperature module, Data Sheet, 2003.
\item[[ 13]]	Freescale Semiconductor, Inc.: MPXH6101A - Integrated Silicon Pressure Sensor for Manifold Absolute Pressure Applications On-Chip Signal Conditioned, Temperature Compensated and Calibrated, Data Sheet, 2009.
\item[[ 14]]	Analog Devices, Inc.: AD22100 Voltage Output Temperature Sensor with Signal Conditioning, Data Sheet, 1998.
\item[[ 15]]	Vaisala: Vaisala BAROCAP® Digital Barometer PTB330, Data Sheet, 2010.
\item[[ 16]]	Ahlborn: ALMEMO® Handbuch, 9. Auflage, Data Sheet, 2011
\item[[ 17]]	Hrabina, J., Lazar, J., Holá, M., Číp, O., Frequency Noise Properties of Lasers for Interferometry in Nanometrology, Sensors 13, 2206-2219 (2013).

\end{itemize}
\end{document}